\documentclass[12pt,a4paper]{article}
\usepackage{graphicx} 
\usepackage[utf8]{inputenc}
\usepackage[T1]{fontenc}
\usepackage{amsmath}  
\usepackage{amsfonts} 
\usepackage{amsthm}
\usepackage{amssymb}
\usepackage{geometry}
\usepackage{booktabs}
\usepackage{threeparttable}
\usepackage{xcolor}
\usepackage{float}
\usepackage[font=small,labelfont=bf]{caption}
\usepackage{array}
\usepackage{multirow}
\usepackage[round,authoryear]{natbib}
\usepackage[colorlinks=true,linkcolor=blue,citecolor=blue,urlcolor=blue]{hyperref}
\definecolor{wbblue}{RGB}{0,114,188}
\makeatletter
\def\@maketitle{%
  \newpage
  \null
  \vskip 2em%
  \begin{center}%
  \let \footnote \thanks
    {\LARGE\bfseries \textit{``Unmatched''} \par}%
    \vskip 1.2em%
    {\LARGE {From Skewed Births to a Structural Surplus of Grooms} \par}%
    \vskip 1.5em%
    {\large
     \lineskip .5em%
      \begin{tabular}[t]{c}%
        Praveen N\footnote{Center of Policy Research and Governance, New Delhi. Email: praveenjai.jk@gmail.com} \\
        Suddhasil Siddhanta\footnote{Gokhale Institute of Politics and Economics, Pune. Email: suddhasil.siddhanta@gipe.ac.in}
      \end{tabular}\par}%
    \vskip 1em%
    {\large January 2026}%
  \end{center}%
  \par
  \vskip 1.5em}
\makeatother

\begin{document}

\maketitle

\begin {abstract}

{Data on marriage flows are not available in most developing countries, making marriage market imbalance difficult to measure. Existing measures use crude fertility rates and do not account for early-life mortality, overstating the number of births surviving to marriageable ages. This paper develops the Surplus Groom Index to quantify marriage market imbalance under monogamy using census age structure, vital registration of births and deaths, and marriage timing data. The index incorporates effective fertility—total births adjusted for under-five mortality—to reflect actual cohort progression from birth to marriageable ages. This adjustment matters in settings where child mortality shapes the supply of marriage partners.
Using India's 2011 Census data, we find that eleven percent of men aged 15–54 cannot marry due to bride shortage, approximately 39 million men. Marriage imbalance is widespread rather than regionally concentrated. Punjab records the highest deficit at 33 percent, but states considered demographically progressive show substantial imbalance: Kerala 18 percent, West Bengal 14 percent, Karnataka and Tamil Nadu 11 percent each. Declining fertility has produced smaller female cohorts unable to absorb male-heavy cohorts from earlier birth years. Balanced sex ratios at birth do not ensure marriage market equilibrium once fertility declines and marriage is delayed.
}

\vspace{0.5em}
\noindent {\textbf{Keywords:}} {Marriage market, masculinity, fertility decline, marriage squeeze, surplus grooms}

\vspace{0.5em}
\noindent {\textbf{JEL Classification:}} {J12, J13, J16}
\end{abstract}

\newpage

\section{Introduction}

India's sex ratio imbalance is well documented. Son preference, manifested through sex-selective abortion and differential care of female children, has produced male-biased sex ratios across much of the country \citep{miller1981, dasgupta2005}. The spread of prenatal sex-determination technology since the 1980s has intensified this pattern, particularly in north-western and western states \citep{guilmoto2007}. But how these imbalances in birth cohorts translate into marriage market outcomes remains poorly measured.

The relationship between sex ratio imbalances and marriage market constraints is not straightforward. Under conditions of high fertility and expanding cohort size, excess men can be matched with women from younger cohorts, particularly where spousal age gaps are flexible \citep{coale1972, dyson1983}. As fertility declines and marriage is delayed, however, the demographic space for such cross-cohort matching narrows \citep{akers1967, dyson2001}, and earlier sex composition imbalances translate into structural deficits of marriageable women \citep{guilmoto2012}. Measuring this transition requires accounting for cohort progression, fertility decline, and marriage timing within an explicit demographic framework \citep{dalbis2012}.

This paper develops the Surplus Groom Index to quantify marriage market imbalance. The index builds on the marriage market framework of \citet{dalbis2012}, which characterizes equilibrium under monogamy through the interaction of fertility, sex ratios at birth, and spousal age gaps. We introduce a refinement that replaces crude fertility with effective fertility---births adjusted for early-life mortality---to reflect actual cohort sizes reaching marriageable ages. This modification matters in settings where child mortality shapes the supply of marriage partners. In India, where under-five mortality averaged 5.2 percent nationally in 2011 and exceeded 7 percent in several states, crude fertility overstates cohort growth and understates marriage market imbalance.

Using 2011 Census and Sample Registration System data for Indian states, we find that marriage imbalance has become widespread rather than regionally concentrated. Nationally, eleven percent of men aged 15--54 will not marry due to bride shortage. In Punjab this reaches thirty-three percent. But the problem extends well beyond north-western states traditionally associated with skewed sex ratios \citep{agnihotri2000}. Kerala shows eighteen percent surplus grooms, Karnataka and Tamil Nadu eleven percent, West Bengal fourteen percent. Even states with relatively balanced sex ratios face substantial marriage deficits. The mechanism is demographic: fertility decline has shrunk younger female cohorts below the size needed to match male-heavy cohorts from earlier birth years. States once considered demographically progressive \citep{dyson1983} now produce structural surplus of marriageable men.

The paper is organized as follows. Section 2 provides background on sex ratio imbalance in India and the motivation for developing a new measure of marriage market outcomes. Section 3 derives the Surplus Groom Index, beginning with the \citet{dalbis2012} framework and introducing the effective fertility modification. Section 4 describes data sources and measurement procedures. Section 5 presents empirical results at the national and state levels. Section 6 discusses the findings and their implications. Section 7 concludes.

\section{Backdrop} 

\subsection{Sex Ratio Imbalance in India}

India exhibits one of the world's most pronounced sex ratio imbalances. \citet{sen1990, sen1992} brought global attention to the phenomenon of ``missing women,'' estimating that approximately 46 million females were missing in India alone \citep{unfpa2020}. This deficit reflects sustained son preference manifested through sex-selective abortion, differential postnatal care, and excess female mortality \citep{dasgupta2005, jha2011}. The child sex ratio (CSR, ages 0-6) has deteriorated sharply since the early 1990s, coinciding with the spread of prenatal sex-determination technology \citep{guilmoto2007}.

Table 1 documents the evolution of child sex ratios across major Indian states from 1961 to 2011. The CSR provides a cumulative measure of sex-selective practices and differential child mortality, with values substantially above the biologically normal range of 103-105 indicating gender discrimination \citep{coale1991}.

\begin{table}[htbp]
\centering
\caption{Trends in Child Sex Ratios (0--6 years), India and Major States, 1961--2011}

\label{tab:csr_states}
\begin{threeparttable}
\begin{tabular}{lccccc}
\toprule
State & 1961 & 1981 & 1991 & 2001 & 2011 \\
\midrule
Odisha & 97 & 101 & 101 & 105 & 106 \\
Bihar & 101 & 102 & 102 & 106 & 107 \\
Tamil Nadu & 102 & 103 & 103 & 106 & 106 \\
Himachal Pradesh & 102 & 103 & 103 & 112 & 110 \\
Madhya Pradesh & 102 & 102 & 102 & 107 & 109 \\
Chhattisgarh &  &  &  & 103 & 103 \\
Maharashtra & 102 & 105 & 105 & 110 & 112 \\
Gujarat & 105 & 106 & 106 & 113 & 112 \\
Rajasthan & 105 & 105 & 105 & 110 & 113 \\
Uttar Pradesh & 106 & 107 & 108 & 109 & 111 \\
Uttarakhand &  &  &  & 110 & 112 \\
Punjab & 110 & 110 & 114 & 125 & 118 \\
Haryana &  & 114 & 114 & 122 & 120 \\
West Bengal & 99 & 102 & 103 & 104 & 105 \\
\bottomrule
\end{tabular}
\begin{tablenotes}[flushleft]
\footnotesize
\item \textit{Source}: 2011 data are taken from census report whereas 1961, 1981, 1991 and 2001 data have been taken from the Siddhanta (2009) PhD thesis, \textit{On Gender Bias in Child Population in India: A Fresh Exploration}.
\end{tablenotes}
\end{threeparttable}
\end{table}

Several patterns emerge. First, north-western states---Punjab, Haryana, Himachal Pradesh, Rajasthan---exhibit persistently elevated CSR values throughout the period, with Punjab reaching 118 by 2011. This regional concentration reflects longstanding kinship patterns and agricultural systems favoring sons \citep{dyson1983, miller1981, agnihotri2000}. Second, western states show deterioration over time. Gujarat increased from 105 in 1961 to 112 in 2011, while Maharashtra rose from 102 to 112. This pattern of ``northernization'' \citep{siddhanta2009} indicates that son preference has intensified in regions historically considered more gender-balanced. Third, even southern states like Tamil Nadu and Karnataka, often cited as demographically progressive, now show CSR values of 106, above the normal range.

A substantial literature documents marked regional disparities in sex ratios across India. Empirical studies consistently identify a pronounced divide between the north-western and south-eastern regions, with the former exhibiting a strong preponderance of male births \citep{sopher1980, miller1981, agnihotri2000}. This divide has been attributed to kinship structures and cultural practices: patrilocal, patrilineal systems in the north-west generate strong son preference, while more bilateral kinship in the south produces more balanced sex ratios \citep{dyson1983}.

However, recent evidence suggests this traditional regional classification has weakened. Sex ratio distortion has spread beyond its traditional core \citep{siddhanta2009, guilmoto2012}, with states historically considered favorable to females---Tamil Nadu, Andhra Pradesh, Karnataka---now showing elevated CSR values (Table 1). This spatial diffusion has important implications for marriage markets, as cohort sex composition imbalances established in childhood persist into adulthood \citep{guilmoto2012}.

The Indian pattern differs notably from China, where sex ratio distortion is commonly linked to fertility restriction under the one-child policy \citep{coale1996}. In India, fertility decline does not automatically produce sex ratio normalization. Instead, in several regions fertility decline has coincided with intensified sex selection, reinforcing rather than correcting cohort imbalance \citep{siddhanta2014, jayachandran2017}. This suggests that son preference remains strong even as family size declines, with parents using sex-selective abortion to achieve desired sex composition within smaller families.

\subsection{Motivation of the Study}

Sex imbalances established at birth persist through cohort progression, generating long-term consequences for marriage markets. But the extent of marriage market imbalance across India remains inadequately measured. Existing research focuses primarily on sex ratios at birth or in early childhood \citep{guilmoto2007, jha2011}, or on regional variation in juvenile sex ratios \citep{agnihotri2000, siddhanta2009}. These indicators capture gender imbalance at early life stages but do not directly quantify constraints within marriageable cohorts, which emerge over time through cohort progression and are shaped by fertility dynamics and marriage timing.

The gap between sex ratio indicators and marriage market outcomes exists because skewed sex ratios do not directly produce marriage squeeze. Under high fertility, excess men in one cohort can be matched with women from larger, younger cohorts \citep{coale1972, casterline1986}. Under low fertility, this adjustment mechanism weakens, and sex ratio imbalances translate more directly into marriage squeeze \citep{anderson2007, guilmoto2012}. The translation from sex ratios to marriage market outcomes thus depends critically on fertility dynamics and marriage timing \citep{dalbis2012}.

India is experiencing rapid demographic transition. Fertility has declined from 5.2 children per woman in 1971 to 2.4 in 2011, while infant mortality fell by two-thirds over the same period \citep{guilmoto2013}. However, demographic transition has not produced sex ratio normalization. Rather, it has coincided with intensified son preference, as smaller family sizes increase the salience of sex composition \citep{dasgupta1997}. Whether this combination of declining fertility and persistent sex ratio imbalance has produced measurable marriage market constraints across India, and how to quantify such constraints at the population level, remains empirically and methodologically unresolved. Existing frameworks for measuring marriage market balance either require detailed data on marriage flows \citep{schoen1983} or rely on crude fertility rates that may overstate cohort growth in high-mortality settings \citep{dalbis2012}. In developing countries where marriage registration is incomplete and under-five mortality remains substantial, neither approach provides a tractable measure of marriage market imbalance.

This paper addresses this gap by developing the Surplus Groom Index, a demographic measure of marriage market imbalance constructed within an age-structured population framework. The index builds on \citet{dalbis2012}, who characterize equilibrium under monogamy through the interaction of fertility, sex ratios at birth, and spousal age gaps. We introduce a targeted refinement that replaces crude fertility with effective fertility---total births adjusted for under-five mortality---to account for early-life demographic attrition directly. This modification preserves the tractability of the original framework while improving measurement accuracy in high-mortality contexts. The index requires only census age structure, vital registration of births and deaths, and marriage age distributions---data routinely collected even in countries lacking marriage registration systems.

\section{The Surplus Groom Index}

The Surplus Groom Index measures marriage market imbalance under monogamy. The index builds on the demographic accounting framework of \citet{dalbis2012}, which characterizes marriage market equilibrium through the interaction of fertility, sex ratios at birth, and spousal age gaps. Their framework provides conditions under which marriage markets clear, using total fertility as a measure of cohort growth. This assumes that all births survive to marriageable ages.

We replace crude fertility with effective fertility---births adjusted for under-five mortality. Let $r_t$ denote the total fertility rate per married woman and $u_t$ the under-five mortality rate. Effective fertility is given by:
\begin{equation}
\tilde{r}_t = r_t(1 - u_t)
\end{equation}

This adjustment incorporates early-life demographic attrition into the population accounting of future marriage cohorts.\footnote{Early-life mortality remains substantial in developing countries and varies across time and regions \citep{preston2001, un2019}. In South Asia, under-five mortality is not gender neutral, reflecting postnatal care biases and differential intra-household resource allocation \citep{sen1990, dasgupta2005, bhalotra2010}. Fertility measured purely as births per woman may overstate the effective supply of future brides relevant for the marriage market.} By defining population growth in terms of effective fertility, the index reflects actual cohort survival to marriageable ages rather than births alone.

Marriage market balance depends on the timing of entry into marriage. Let $A_m$ and $A_f$ denote the average ages at marriage of men and women, respectively, and let $\alpha$ denote the average age at first birth.\footnote{See Appendix B for details on the construction of age at marriage using the SMAM method.} The spousal age gap $\Delta A = A_m - A_f$ governs the extent to which men can match with women from younger cohorts. Larger age gaps facilitate inter-cohort matching, while declining fertility and delayed marriage compress the demographic window over which this adjustment can operate.

Let $S_t$ denote the sex ratio at birth, defined as the number of females per male. The Surplus Groom Index at time $t$ is:\footnote{The formal derivation of the Surplus Groom Index is provided in Appendix A.}
\begin{equation}
\text{SGI}_t = \frac{1}{S_t} \left[\frac{1 + S_t}{\tilde{r}_t S_t}\right]^{\frac{\Delta A}{A_f + \alpha}}
\end{equation}

When demographic growth and marriage timing are sufficient to offset sex imbalances at birth, the marriage market clears under monogamy. When these conditions fail to hold, the number of men reaching marriageable age exceeds the number of women available for matching, resulting in a structural surplus of grooms. The index can be interpreted as the proportion of men unable to marry due to demographic constraints. For example, $\text{SGI} = 1.11$ indicates that 11 percent of men in marriageable cohorts will remain unmarried.

\section{Data and Measurement}

The empirical analysis uses data from the 2011 Population Census of India and the Sample Registration System (SRS). The Census provides complete information on population size, age structure, marital status, and sex composition at the state level. The unit of analysis is the major Indian state, defined according to administrative boundaries prevailing at the time of the 2011 census. We use census tabulations to obtain sex ratios at birth, population counts by single years of age, and marital status distributions by age and sex. These data form the basis for constructing the demographic cohorts relevant to the marriage market. The sex ratio at birth is defined as the number of females per male.

Information on marriage timing is drawn from census-based age distributions. Average ages at marriage for men and women are estimated using the Singulate Mean Age at Marriage (SMAM) method originally proposed by \citet{hajnal1953}, which remains the standard approach in census-based demographic analysis. SMAM estimates the mean age at first marriage among those who eventually marry, based on the proportion never married at each age.

Fertility and mortality indicators are drawn from the Sample Registration System, India's principal source of subnational vital statistics. State-level total fertility rates are taken from SRS estimates for the period 2001–03, chosen to align completed fertility with the cohorts observed in the 2011 census. Under-five mortality rates are obtained from the 2011–12 SRS life tables and are used to adjust fertility to account for early-life demographic attrition between birth and marriageable ages. For smaller states where under-five mortality estimates are unavailable—most notably Mizoram and Nagaland—infant mortality rates from the 2017 SRS are used as a proxy.

By combining census-based cohort structures with fertility and mortality estimates from the SRS, the analysis constructs an effective fertility measure that reflects the number of births surviving early childhood rather than total births alone. This effective fertility rate is calculated as total fertility rate multiplied by one minus the under-five mortality rate. All variables used in the computation of the Surplus Groom Index are drawn from official demographic sources and harmonized to ensure consistency in timing and population coverage.

\section{Result}

The Surplus Groom Index shows that marriage imbalance is no longer confined to the north-western states traditionally associated with skewed sex ratios, but is observable across India. Figure 1 presents the kernel density distribution of SGI values across Indian states. As can be seen from the density plot, the distribution is clearly shifted to the right of the equilibrium benchmark of unity (shown as the green vertical line). The peak of the distribution lies around 1.11 (shown as the black vertical line), which represents the national average. Most states cluster around this value, indicating that approximately 11 percent excess men in marriageable ages is the typical pattern across India. Most states cluster above unity, and the density exhibits a long upper tail, indicating that substantial excess male cohorts are present in many parts of the country. The concentration of states above the threshold confirms that current demographic conditions are no longer sufficient to offset male-heavy cohort structures through fertility or marriage timing alone. What emerges is not a narrow or marginal squeeze, but a broad structural imbalance affecting a large share of the population.

\begin{figure}[htbp]
    \centering
    \includegraphics[width=0.8\textwidth]{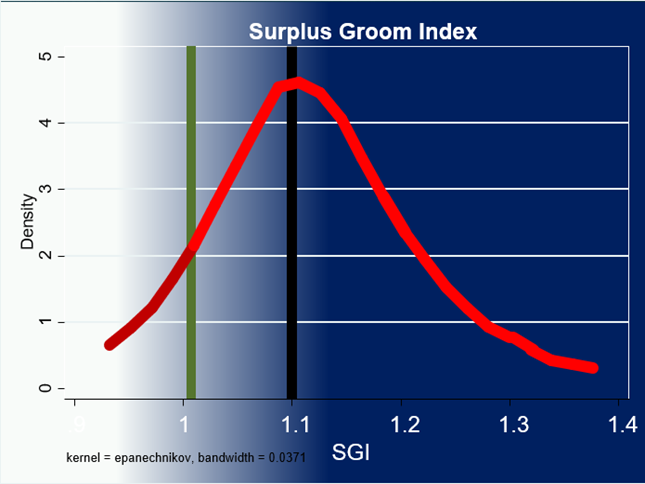}
    \caption{Surplus Groom Index: All India (K-density plot)}
    \label{fig:k_density_plot}
    \vspace{1ex}

\end{figure}

Table 2 reports SGI values for major Indian states. At the national level, the SGI equals 1.11, indicating that 11 percent of men aged 15–54 will not find brides under prevailing demographic conditions. This corresponds to approximately 39 million men based on the 2011 Census population. The state-level data show considerable variation. While Punjab records the highest SGI of 1.33, Meghalaya shows the lowest value of 0.97. However, most states fall between 1.05 and 1.20, suggesting that marriage imbalance has become a general feature rather than an anomaly confined to specific regions.

\begin{table}[h!]
\centering
\begin{threeparttable}
\caption{Surplus Groom Index: Major States 1991 -- 2011}
\begin{tabular}{l c}
\toprule
\textbf{States} & \textbf{Surplus Groom Index (2011)} \\
\midrule
Punjab & 1.33 \\
Jammu and Kashmir & 1.26 \\
Himachal Pradesh & 1.23 \\
Haryana & 1.21 \\
Kerala & 1.18 \\
Maharashtra & 1.18 \\
Gujarat & 1.15 \\
Uttarakhand & 1.15 \\
West Bengal & 1.14 \\
Manipur & 1.13 \\
Karnataka & 1.11 \\
Rajasthan & 1.11 \\
Andhra Pradesh & 1.11 \\
Tamil Nadu & 1.11 \\
Uttar Pradesh & 1.11 \\
Nagaland & 1.10 \\
Jharkhand & 1.09 \\
Mizoram & 1.07 \\
Assam & 1.07 \\
Bihar & 1.07 \\
Orrisa & 1.06 \\
Madhya Pradesh & 1.03 \\
Chhattisgarh & 0.99 \\
Meghalaya & 0.97 \\
India & 1.11 \\
\bottomrule
\end{tabular}
\begin{tablenotes}
\footnotesize
\item Note: Author’s computation
\end{tablenotes}
\end{threeparttable}
\end{table}

Several salient features are forthcoming from Table 2. First, the highest levels of imbalance are not just confined to the north-western states, though this region clearly shows the most severe constraints. Punjab stands out with an SGI of 1.33, meaning that about one out of every three men in marriageable ages cannot find a bride. Jammu and Kashmir (1.26), Himachal Pradesh (1.23), and Haryana (1.21) also record substantially high SGI values. Gujarat (1.15) and Uttarakhand (1.15) complete the north-western cluster. These states have long been characterised by male-biased sex ratios at birth; the SGI shows that, with fertility now falling to low levels, these imbalances translate directly into shortfalls of women in the marriage market.

Second, southern states that were historically considered demographically balanced now display substantial marriage market imbalance. Kerala records an SGI of 1.18, implying that nearly one in five men in marriageable ages faces a shortage of women, despite relatively favourable sex ratios and high female literacy. Similarly, Karnataka (1.11), Tamil Nadu (1.11), and Andhra Pradesh (1.11) all show SGI values at or above the national average. Interestingly, West Bengal in the east also records a relatively high value of 1.14. These results indicate that balanced sex ratios at birth do not ensure balance in the marriage market once fertility declines and marriage is delayed.

Third, geographically large and densely populated states contribute heavily to the absolute number of surplus men even when their SGI values are closer to the national average. Uttar Pradesh (1.11), Maharashtra (1.18), and Rajasthan (1.11) each accounts for several million excess men due to their population size. Maharashtra alone contributes more than six million, while Uttar Pradesh adds a similar number. Bihar, despite an SGI of only 1.07, contributes substantially to the total owing to its huge population base. The scale of these populations means that even moderate imbalance results in large absolute deficits of marriageable women.

Fourth, only two states—Meghalaya (0.97) and Chhattisgarh (0.99)—show SGI values at or below unity, where the number of women in marriageable ages slightly exceeds that of men. These states represent exceptions to the general pattern and remain spatially isolated rather than forming part of larger regional clusters.

\subsection{Spatial Patterns and Regional Clustering}

Figure 2 presents the spatial distribution of SGI values across Indian states. The map reveals clear regional contiguities rather than isolated state-level anomalies. As can be seen from the map, high levels of marriage imbalance form a contiguous belt across north-western India, extending from Punjab and Haryana through Rajasthan and into Gujarat and Uttarakhand. This belt corresponds to regions long characterized by male-biased sex ratios, but the SGI indicates that these distortions now translate into sustained marriage imbalance across adjoining states rather than remaining confined to individual administrative units.

\begin{figure}[htbp]
    \centering
    \includegraphics[width=0.8\textwidth]{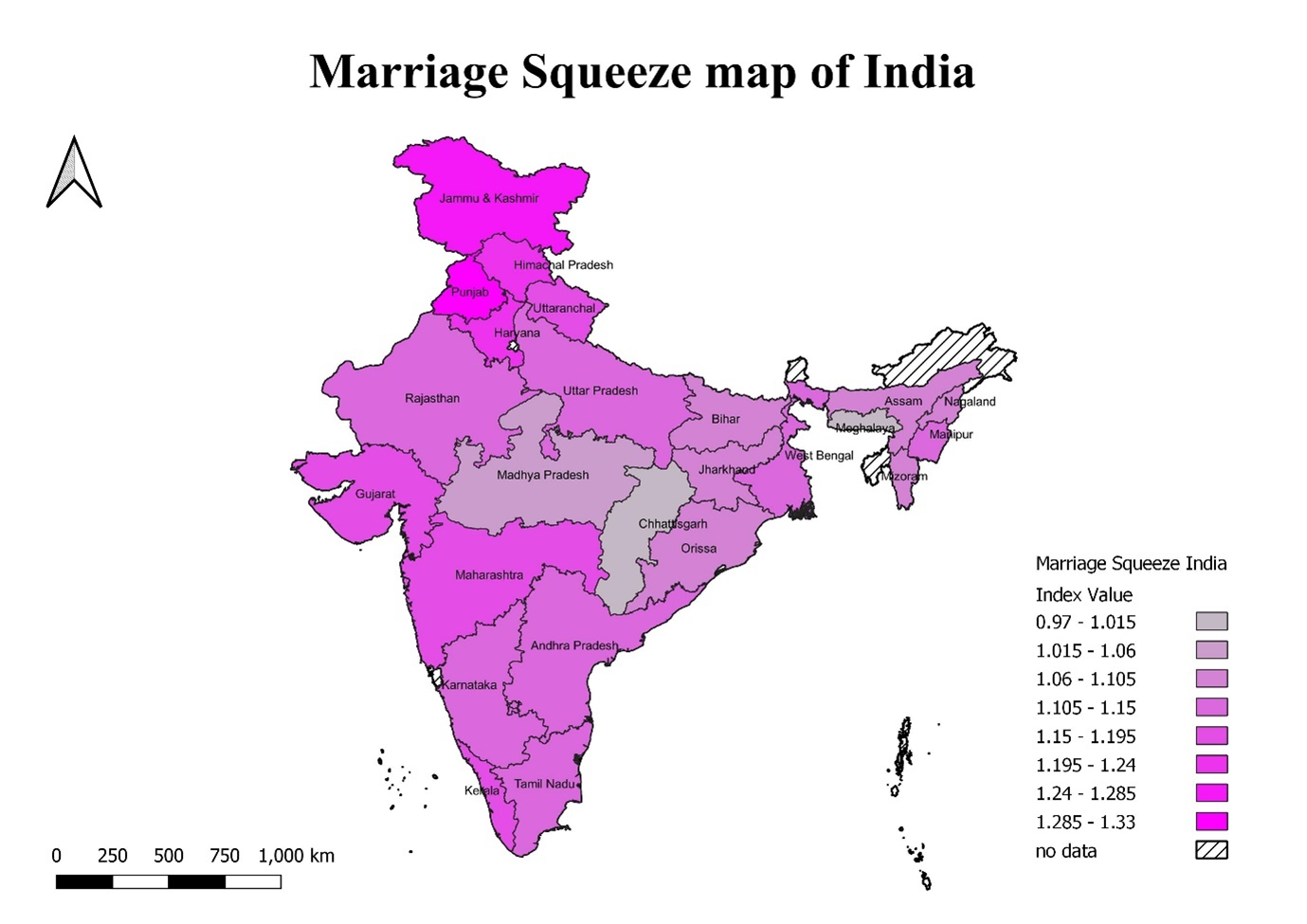}
    \caption{Marriage Squeeze Index, All India, State Level, 2011}
    \label{fig:marriage_squeeze_map}
    \vspace{1ex}
    \footnotesize Note: Author’s computation
\end{figure}

A second cluster can be observed in western and southern India, where states such as Maharashtra, Karnataka, Tamil Nadu, and Kerala—despite distinct social and demographic histories—display similarly elevated SGI values (all between 1.11 and 1.18). The appearance of this southern cluster is particularly noteworthy. These states were historically characterized by relatively balanced sex ratios and more progressive social indicators. Their emergence as marriage-constrained regions seems to reflect the compression of cohort size under fertility decline rather than extreme sex ratio distortion at birth. In eastern India, West Bengal and adjoining states also register imbalance above the equilibrium threshold, though the values are moderate (around 1.05-1.15).

Close inspection of the map points out that the few states with SGI values at or below unity—notably Meghalaya and Chhattisgarh—remain spatially isolated. The map also shows that surplus male cohorts are no longer restricted to a single geographic zone. The spatial clustering suggests that marriage market constraints might operate through regional demographic processes that transcend administrative boundaries.

The traditional 'North-South' divide in demographic outcomes—rooted in kinship systems, inheritance practices, and female autonomy \citep{dyson1983, sopher1980}—no longer maps neatly onto marriage market patterns. Southern states with historically favourable conditions for women now face marriage constraints comparable to some northern states, though the underlying demographic mechanisms seem to differ. In the north-west, imbalance stems from persistently skewed sex ratios at birth combined with fertility decline. In the south, imbalance appears to emerge primarily from the compression of cohort size under low fertility, even where sex ratios remain relatively balanced.

Interestingly, the coastal pattern visible in fertility decline—where coastal districts transitioned earlier than interior regions—does not correspond to marriage market outcomes. Coastal states like Kerala, Tamil Nadu, and Maharashtra exhibit substantial marriage imbalance despite having completed fertility transition decades ago. This suggests that early fertility decline, rather than protecting against marriage squeeze, may have contributed to current imbalance by compressing cohort sizes before sex ratio distortions could be corrected.

The emergence of imbalance in economically advanced states indicates that development and demographic modernisation do not automatically resolve marriage market constraints. States like Kerala, Tamil Nadu, Maharashtra, and Gujarat have achieved high levels of human development, yet all record SGI values between 1.11 and 1.18. The spatial pattern suggests that marriage market imbalance in India might reflect structural demographic arithmetic—the interaction of fertility decline, sex ratio bias, and marriage timing—rather than simply underdevelopment or regional backwardness.

The inter-state variation in SGI values ranges from 0.97 to 1.33, with most states clustered between 1.05 and 1.20. This relatively narrow range around the national average, combined with the broad spatial coverage of elevated SGI values, indicates that marriage imbalance has become a general feature of India's demographic landscape. The implications extend beyond individual marriage prospects to encompass household formation patterns, migration dynamics, and potentially labor market and savings behavior across substantial portions of the adult male population in coming decades.

\section{Discussion}

Demographic theory has long emphasized that skewed sex ratios do not mechanically produce marriage squeeze. Under conditions of high fertility and expanding cohort size, excess men in one cohort may be matched with women from younger cohorts, particularly where spousal age gaps are flexible \citep{coale1972, schoen1983, dyson1983, casterline1986}. The SGI results indicate that these adjustment mechanisms have weakened substantially. With fertility at or below replacement in many states and marriage increasingly delayed, the demographic space for cross-cohort matching has narrowed, allowing earlier sex composition imbalances to translate directly into shortfalls of women at marriageable ages.

This process is particularly visible in economically advanced states. Punjab, Haryana, Gujarat, Maharashtra, Kerala, and Tamil Nadu combine low fertility, rising ages at marriage, and elevated SGI values. Close inspection of the results reveals that marriage imbalance in these settings reflects the compression of cohort size under fertility decline rather than current sex ratios alone. As smaller female cohorts enter marriageable ages, they can no longer absorb male-heavy cohorts shaped by earlier demographic conditions \citep{guilmoto2012, eklund2013}. High SGI values in prosperous states thus reflect the arithmetic of demographic transition rather than regional anomaly.

Interestingly, states long considered demographically progressive now face marriage market constraints. Kerala, with high female literacy and relatively balanced sex ratios, records an SGI of 1.18. Tamil Nadu (1.11), Karnataka (1.11), and Andhra Pradesh (1.11) all show substantial imbalance. These findings indicate that balanced sex ratios at birth do not ensure equilibrium in the marriage market once fertility declines and marriage is delayed. The coastal pattern visible in fertility decline does not map neatly onto marriage market outcomes. Early fertility decline may have contributed to current imbalance by compressing cohort sizes before sex ratio distortions could be corrected.

The spatial pattern also points out important regional dynamics. The contiguous belt of high imbalance across north-western India corresponds to regions long characterized by male-biased sex ratios. However, the emergence of a second cluster in western and southern India indicates that marriage imbalance now operates through multiple demographic mechanisms. In the north-west, imbalance stems from persistently skewed sex ratios at birth combined with fertility decline. In the south, imbalance appears to emerge primarily from cohort compression under low fertility, even where sex ratios remain relatively balanced. The traditional 'North-South' divide in demographic outcomes \citep{dyson1983, sopher1980} no longer provides an adequate framework for understanding contemporary marriage market patterns.

Sex ratios describe numerical balance at specific ages but do not capture how demographic imbalance unfolds as cohorts progress through the life course. The Surplus Groom Index addresses this limitation by translating population structure into a measure of demographic matchability under monogamous marriage. By incorporating effective fertility, cohort growth, and spousal age gaps, the SGI captures the cumulative effects of past demographic processes rather than isolated outcomes at birth or childhood. An important implication is that balanced sex ratios do not guarantee marriage market equilibrium. Several states with relatively favourable sex ratios nonetheless exhibit substantial marriage imbalance once fertility declines and marriage is delayed. Previous research has shown that stabilization of sex ratios does not reverse marriage squeeze once cohort effects are locked in by low fertility \citep{guilmoto2007, guilmoto2012}, and formal models demonstrate that imbalance can persist even as sex ratios approach parity when cohort growth slows \citep{dalbis2012}. The SGI results provide quantitative evidence of this mechanism operating broadly across Indian states.

The modification introduced in this paper—replacing crude fertility with effective fertility adjusted for under-five mortality—represents an important refinement. In developing country settings where early-life mortality remains substantial and exhibits gender differentials, crude fertility overstates the effective supply of future brides and grooms. This modification seems particularly important in the Indian context, where under-five mortality rates vary substantially across states and gender differentials in early-life mortality have been documented \citep{sen1990, dasgupta2005, bhalotra2010}. By expressing imbalance both as an index and as a proportion of men affected, the SGI allows comparison across states with very different population sizes. Uttar Pradesh, Maharashtra, and Bihar—despite SGI values between 1.07 and 1.18—together contribute more than fifteen million surplus men due to their population size.

The scale and geographic spread of marriage imbalance suggest important implications for social and economic outcomes. With approximately 39 million men aged 15-54 facing a structural shortage of marriageable women, adaptive responses are likely to emerge. Previous research documents cross-regional marriage migration increasing substantially, with men from high-imbalance states seeking brides from distant regions \citep{kaur2016, blanchet2005}. Delayed marriage represents another likely response. \citet{bhat1999} document rising age at marriage in several north Indian states. Economic responses documented in other contexts—such as increased savings rates to improve marriage prospects \citep{wei2011}—may operate in India, potentially affecting household behavior and asset accumulation patterns. However, the SGI measures demographic constraint rather than predicting individual outcomes. Adaptive responses, regional migration, and changes in marriage norms may partially mitigate demographic constraints. The index provides a benchmark for understanding marriage market pressure rather than a deterministic forecast.

\section{Conclusion and limitations}

This paper develops the Surplus Groom Index to quantify marriage market imbalance in India. The index refines the demographic framework of \citet{dalbis2012} by incorporating effective fertility—births adjusted for early-life mortality—to reflect actual cohort sizes reaching marriageable ages. The 2011 Census data show that marriage market constraints have become widespread rather than regionally concentrated. Nationally, eleven percent of men aged 15–54 will remain unmarried due to bride shortage. Punjab records the highest imbalance (SGI = 1.33), but substantial deficits appear even in states historically considered demographically balanced—Kerala (1.18), Karnataka (1.11), Tamil Nadu (1.11), and West Bengal (1.14). The traditional North-South divide in demographic outcomes no longer provides an adequate framework for understanding contemporary marriage market patterns.

The mechanism is straightforward. Fertility decline has compressed younger female cohorts below the size needed to match male-heavy cohorts shaped by earlier sex composition imbalances. Under high fertility, cross-cohort matching could absorb excess men through flexible spousal age gaps. As fertility approaches replacement and marriage is delayed, this adjustment mechanism weakens, allowing earlier sex ratio distortions to translate directly into marriage market deficits. States that completed fertility transition early—Kerala, Tamil Nadu, Maharashtra—now face marriage constraints precisely because cohort compression occurred before sex ratio normalization. Balanced sex ratios at birth do not guarantee balanced marriage markets once fertility declines. Demographic modernization does not automatically resolve marriage market constraints.

The analysis has important limitations. The index assumes monogamy and universal marriage, which may not hold uniformly across all Indian contexts. It does not account for marriage market segmentation by caste, religion, education, or region—factors that could intensify constraints within specific demographic subgroups even as aggregate imbalance persists. The analysis uses state-level data, which hides variability within states, particularly in geographically large and demographically diverse regions. The effective fertility adjustment relies on under-five mortality rates, but gender differentials in child mortality vary across states and may not be fully captured in aggregate mortality measures. Finally, the index measures demographic constraint rather than predicting individual outcomes. Adaptive responses—migration, delayed marriage, changes in marriage norms—may partially mitigate demographic pressures in ways not captured by the demographic accounting framework.

Future research can extend the framework to account for educational homogamy, caste endogamy, or regional marriage market segmentation. How men and women adapt to marriage market constraints—through cross-regional marriage migration \citep{kaur2004, blanchet2005}, delayed marriage, hypergamy adjustments, or changes in household formation patterns—remains an important question for investigation. The implications of surplus male cohorts for gender relations, intra-household bargaining, and female autonomy also need empirical attention. The SGI can be applied to other settings experiencing concurrent fertility decline and sex ratio distortion, particularly in East and Southeast Asia where similar demographic conditions prevail.

What emerges from this analysis is that demographic transition in India has produced unintended consequences for marriage markets. Fertility decline has interacted with sex ratio distortion to create structural imbalance affecting tens of millions of people. The scale and geographic spread of marriage market imbalance carries broader implications beyond individual marriage prospects. Research on China and other Asian societies documents that surplus male populations may affect social stability \citep{hudson2004}, increase savings rates as families compete to improve marriage prospects \citep{wei2011}, and trigger cross-regional marriage migration \citep{kaur2004, blanchet2005}. The marriage market effects of past imbalances will persist for decades as distorted cohorts move through marriageable ages \citep{tuljapurkar1995, jiang2013, guilmoto2012}. Policy responses must operate on multiple temporal scales: preventing future imbalance through sex ratio normalization while managing the consequences of existing cohort distortions. The demographic arithmetic documented here suggests that marriage market imbalance will remain a significant feature of Indian demography well into the coming decades.

\newpage
\bibliography{references}

@book{agnihotri2000,
  title={Sex Ratio Patterns in the Indian Population: A Fresh Exploration},
  author={Agnihotri, Satish B.},
  year={2000},
  publisher={Sage Publications},
  address={New Delhi}
}

@article{akers1967,
  title={On measuring the marriage squeeze},
  author={Akers, Donald S.},
  journal={Demography},
  volume={4},
  number={2},
  pages={907--924},
  year={1967}
}

@article{anderson2007,
  title={The economics of dowry and brideprice},
  author={Anderson, Siwan},
  journal={Journal of Economic Perspectives},
  volume={21},
  number={4},
  pages={151--174},
  year={2007}
}

@article{bhat1999,
  title={Demography of brideprice and dowry: Causes and consequences of the Indian marriage squeeze},
  author={Bhat, P. N. Mari and Halli, S. S.},
  journal={Population Studies},
  volume={53},
  number={2},
  pages={129--148},
  year={1999}
}

@techreport{bhalotra2010,
  title={Where have all the young girls gone? Identification of sex selection in India},
  author={Bhalotra, Sonia and Cochrane, Tom},
  type={IZA Discussion Paper},
  number={5381},
  institution={Institute for the Study of Labor (IZA)},
  year={2010}
}

@article{blanchet2005,
  title={Bangladeshi girls sold as brides in North India},
  author={Blanchet, Thérèse},
  journal={Indian Journal of Gender Studies},
  volume={12},
  number={2-3},
  pages={305--334},
  year={2005}
}

@article{bongaarts2015,
  title={How many more missing women? Excess female mortality and prenatal sex selection, 1970-2050},
  author={Bongaarts, John and Guilmoto, Christophe Z.},
  journal={Population and Development Review},
  volume={41},
  number={2},
  pages={241--269},
  year={2015}
}

@article{casterline1986,
  title={The age difference between spouses: Variations among developing countries},
  author={Casterline, John B. and Williams, Lindy and McDonald, Peter},
  journal={Population Studies},
  volume={40},
  number={3},
  pages={353--374},
  year={1986}
}

@book{coale1972,
  title={The Growth and Structure of Human Populations: A Mathematical Investigation},
  author={Coale, Ansley J.},
  year={1972},
  publisher={Princeton University Press}
}

@article{coale1991,
  title={Excess female mortality and the balance of the sexes in the population: An estimate of the number of "missing females"},
  author={Coale, Ansley J.},
  journal={Population and Development Review},
  volume={17},
  number={3},
  pages={517--523},
  year={1991}
}

@article{coale1996,
  title={Five decades of missing females in China},
  author={Coale, Ansley J. and Banister, Judith},
  journal={Proceedings of the American Philosophical Society},
  volume={140},
  number={4},
  pages={421--450},
  year={1996}
}

@article{dalbis2012,
  title={Missing daughters, missing brides?},
  author={d'Albis, Hippolyte and de la Croix, David},
  journal={Economics Letters},
  volume={116},
  number={3},
  pages={358--360},
  year={2012},
  publisher={Elsevier}
}

@article{dasgupta2005,
  title={Explaining Asia's "missing women": A new look at the data},
  author={Das Gupta, Monica},
  journal={Population and Development Review},
  volume={31},
  number={3},
  pages={529--535},
  year={2005}
}

@article{dasgupta1997,
  title={Fertility decline and increased manifestation of sex bias in India},
  author={Das Gupta, Monica and Bhat, P. N. Mari},
  journal={Population Studies},
  volume={51},
  number={3},
  pages={307--315},
  year={1997}
}

@article{dyson2001,
  title={The preliminary demography of the 2001 census of India},
  author={Dyson, Tim},
  journal={Population and Development Review},
  volume={27},
  number={2},
  pages={341--356},
  year={2001}
}

@article{dyson1983,
  title={On kinship structure, female autonomy, and demographic behavior in India},
  author={Dyson, Tim and Moore, Mick},
  journal={Population and Development Review},
  volume={9},
  number={1},
  pages={35--60},
  year={1983}
}

@article{eklund2013,
  title={Marriage squeeze and mate selection: Analysing the ecology of choice and implications for social policy in China},
  author={Eklund, Lisa},
  journal={Economic and Political Weekly},
  volume={48},
  number={35},
  pages={48--55},
  year={2013},
  month={August}
}

@inproceedings{guilmoto2007,
  title={Sex-ratio imbalance in Asia: Trends, consequences and policy responses},
  author={Guilmoto, Christophe Z.},
  booktitle={UNFPA Fourth Asia Pacific Conference on Reproductive and Sexual Health and Rights},
  year={2007}
}

@article{guilmoto2012,
  title={Skewed sex ratios at birth and future marriage squeeze in China and India, 2005-2100},
  author={Guilmoto, Christophe Z.},
  journal={Demography},
  volume={49},
  number={1},
  pages={77--100},
  year={2012}
}

@article{guilmoto2013,
  title={Fertility at the district level in India: Lessons from the 2011 census},
  author={Guilmoto, Christophe Z. and Rajan, S. Irudaya},
  journal={Economic and Political Weekly},
  volume={48},
  number={23},
  pages={59--70},
  year={2013}
}

@article{hajnal1953,
  title={Age at marriage and proportions marrying},
  author={Hajnal, John},
  journal={Population Studies},
  volume={7},
  number={2},
  pages={111--136},
  year={1953}
}

@book{hudson2004,
  title={Bare Branches: The Security Implications of Asia's Surplus Male Population},
  author={Hudson, Valerie M. and den Boer, Andrea M.},
  year={2004},
  publisher={MIT Press}
}

@article{jayachandran2017,
  title={Fertility decline and missing women},
  author={Jayachandran, Seema},
  journal={American Economic Journal: Applied Economics},
  volume={9},
  number={1},
  pages={118--139},
  year={2017}
}

@article{jha2011,
  title={Trends in selective abortions of girls in India: Analysis of nationally representative birth histories from 1990 to 2005 and census data from 1991 to 2011},
  author={Jha, Prabhat and Kesler, Maya A. and Kumar, Rajesh and Ram, Faujdar and Ram, Usha and Aleksandrowicz, Lukasz and Bassani, Diego G. and Chandra, Shailaja and Banthia, J. K.},
  journal={The Lancet},
  volume={377},
  number={9781},
  pages={1921--1928},
  year={2011}
}

@article{jiang2013,
  title={Marriage squeeze, never-married proportion, and mean age at first marriage in China},
  author={Jiang, Quanbao and Feldman, Marcus W. and Li, Shuzhuo},
  journal={Population Research and Policy Review},
  volume={33},
  number={2},
  pages={189--204},
  year={2013}
}

@article{kaur2004,
  title={Across-region marriages: Poverty, female migration and the sex ratio},
  author={Kaur, Ravinder},
  journal={Economic and Political Weekly},
  volume={39},
  number={25},
  pages={2595--2603},
  year={2004}
}

@article{kaur2016,
  title={Mapping the adverse consequences of sex selection and gender imbalance in India and China},
  author={Kaur, Ravinder},
  journal={Economic and Political Weekly},
  volume={51},
  number={35},
  pages={37--44},
  year={2016}
}

@book{miller1981,
  title={The Endangered Sex: Neglect of Female Children in Rural North India},
  author={Miller, Barbara D.},
  year={1981},
  publisher={Cornell University Press}
}

@book{preston2001,
  title={Demography: Measuring and Modeling Population Processes},
  author={Preston, Samuel H. and Heuveline, Patrick and Guillot, Michel},
  year={2001},
  publisher={Blackwell Publishers}
}

@article{schoen1983,
  title={Measuring the tightness of a marriage squeeze},
  author={Schoen, Robert},
  journal={Demography},
  volume={20},
  number={1},
  pages={61--78},
  year={1983}
}

@article{sen1990,
  title={More than 100 million women are missing},
  author={Sen, Amartya K.},
  journal={New York Review of Books},
  volume={37},
  number={20},
  pages={61--66},
  year={1990}
}

@article{sen1992,
  title={Missing women},
  author={Sen, Amartya K.},
  journal={British Medical Journal},
  volume={304},
  number={6827},
  pages={587--588},
  year={1992}
}

@phdthesis{siddhanta2009,
  title={On Gender Bias in Child Population in India: A Fresh Exploration},
  author={Siddhanta, Suddhasil},
  year={2009},
  school={University of Kalyani, Kalyani, West Bengal}
}

@article{siddhanta2014,
  title={Hundred years of juvenile masculinity in India: Why the contemporary pattern is important?},
  author={Siddhanta, Suddhasil and Nandy, Debasish},
  journal={Available at SSRN},
  year={2014},
  note={SSRN Working Paper},
  url={https://ssrn.com/abstract=2535548}
}

@incollection{sopher1980,
  title={Sex disparity in Indian literacy},
  author={Sopher, David E.},
  booktitle={An Exploration of India: Geographical Perspectives on Society and Culture},
  editor={Sopher, David E.},
  pages={129--158},
  year={1980},
  publisher={Cornell University Press}
}

@article{tuljapurkar1995,
  title={High sex ratios in China's future},
  author={Tuljapurkar, Shripad and Li, Nan and Feldman, Marcus W.},
  journal={Science},
  volume={267},
  number={5199},
  pages={874--876},
  year={1995}
}

@report{unfpa2020,
  title={State of World Population 2020},
  author={{UNFPA}},
  year={2020},
  institution={United Nations Population Fund}
}

@report{un2019,
  title={World Mortality Report 2019},
  author={{United Nations}},
  year={2019},
  institution={United Nations Department of Economic and Social Affairs, Population Division}
}

@article{wei2011,
  title={The competitive saving motive: Evidence from rising sex ratios and savings rates in China},
  author={Wei, Shang-Jin and Zhang, Xiaobo},
  journal={Journal of Political Economy},
  volume={119},
  number={3},
  pages={511--564},
  year={2011}
}

\newpage

\appendix

\section{Supplementary Demographic Evidence}

\setcounter{table}{0}
\renewcommand{\thetable}{A\arabic{table}}
\begin{table}[htbp]
\centering
\caption{Estimates of the Number of Missing Females Worldwide}
\label{tab:A1_missing_females}
\begin{threeparttable}
\begin{tabular}{lcc}
\toprule
Author & Number (millions) & Reference Year \\
\midrule
Sen (1989, 1990) & $>$100 & 1990 \\
Coale (1991) & 60 & 1990 \\
Sen (1992) & $>$100 & 1990 \\
Klasen (1994) & 92 & 1990 \\
Klasen and Wink (2002) & 102 & 1990 \\
Klasen and Wink (2003) & 101 & 2000 \\
Guilmoto (2012) & 117 & 2010 \\
Bongaarts and Guilmoto (2015) & 126 & 2010 \\
\bottomrule
\end{tabular}
\begin{tablenotes}[flushleft]
\footnotesize
\item \textit{Notes}: Differences across these estimates primarily reflect methodological choices. Coale (1991) used West model stable populations as the reference, whereas Klasen (1994) and Klasen and Wink (2002, 2003) relied on East model populations and allowed for variation in the sex ratio at birth. Bongaarts and Guilmoto (2015) instead used age distributions from populations without gender discrimination and estimated missing females by age group.
\item \textit{Source}: \citet{bongaarts2015}.
\end{tablenotes}
\end{threeparttable}
\end{table}

\section{Derivation of the Surplus Groom Index}

This appendix derives the Surplus Groom Index (SGI) within an age-structured marriage market framework. While the marriage imbalance condition itself follows directly from cohort matching as in \citet{dalbis2012}, the implicit assumption that fertility maps one-to-one into marriageable cohort size is relaxed by defining population growth in terms of effective fertility, adjusted for early-life mortality.

\subsection{Population structure and births}
Time is continuous. Let $F_t$ and $M_t$ denote the number of female and male births at time $t$, respectively. The sex ratio at birth is defined as
\[
S_t \equiv \frac{F_t}{M_t}.
\]

Let $r_t$ denote total fertility per married woman, and let $u_t$ denote the under-five mortality rate. Effective fertility is defined as
\[
\tilde{r}_t \equiv r_t(1-u_t).
\]

Births at time $t$ are generated by women who entered marriage at time $t-A_f-\alpha$, where $A_f$ denotes the average age at marriage of women and $\alpha$ denotes the average time interval from marriage to first birth, such that women give birth at average age $A_f + \alpha$. Under the assumption of monogamy and universal marriage, total births satisfy
\begin{equation}
B_t = \tilde{r}_t F_{t-A_f-\alpha}. \tag{A1}
\end{equation}

Female births constitute a fraction of total births determined by the sex ratio at birth:
\begin{equation}
F_t = \frac{S_t}{1+S_t} \tilde{r}_t F_{t-A_f-\alpha}. \tag{A2}
\end{equation}

\subsection{Demographic Growth}

Assume that population cohorts grow at a constant rate $n$. Cohort sizes therefore satisfy
\begin{equation}
F_{t-\tau} = F_t e^{-n\tau}, \quad \tau > 0. \tag{A3}
\end{equation}

Applying (A3) to (A2) with $\tau = A_f+\alpha$ yields
\[
F_t = \frac{S_t}{1+S_t} \tilde{r}_t F_t e^{-n(A_f+\alpha)}.
\]

Dividing both sides by $F_t$ and taking natural logarithms gives
\[
0 = \ln\left(\frac{S_t}{1+S_t}\right) + \ln(\tilde{r}_t) - n(A_f+\alpha).
\]

Solving for the demographic growth rate,
\begin{equation}
n = \frac{\ln(\tilde{r}_t) + \ln\left(\frac{S_t}{1+S_t}\right)}{A_f + \alpha}. \tag{A4}
\end{equation}

\subsection{Marriage Market Imbalance}

Let $A_m$ and $A_f$ denote the average ages at marriage of men and women, respectively, and define the spousal age gap as
\[
\Delta A \equiv A_m - A_f.
\]

Under monogamy and universal marriage, marriage market imbalance arises when the number of men reaching marriageable age exceeds the number of women available for matching. This condition can be expressed as
\begin{equation}
n \Delta A + \ln(S_t) \le 0. \tag{A5}
\end{equation}

This condition ensures that the inflow of women into the marriage market is sufficient to match the larger male cohorts implied by the spousal age gap under exponential population growth.

\subsection{Derivation of the Surplus Groom Condition}

Substituting (A4) into (A5) yields
\[
\frac{\ln(\tilde{r}_t) + \ln\left(\frac{S_t}{1+S_t}\right)}{A_f + \alpha} \cdot \Delta A + \ln(S_t) \le 0.
\]

Rearranging:
\[
\frac{\Delta A}{A_f + \alpha}\left[\ln(\tilde{r}_t) + \ln\left(\frac{S_t}{1+S_t}\right)\right] + \ln(S_t) \le 0.
\]

Expanding the second term using $\ln\left(\frac{S_t}{1+S_t}\right) = \ln(S_t) - \ln(1+S_t)$:
\[
\frac{\Delta A}{A_f + \alpha}\ln(\tilde{r}_t) + \frac{\Delta A}{A_f + \alpha}[\ln(S_t) - \ln(1+S_t)] + \ln(S_t) \le 0.
\]

Distributing:
\[
\frac{\Delta A}{A_f + \alpha}\ln(\tilde{r}_t) + \frac{\Delta A}{A_f + \alpha}\ln(S_t) - \frac{\Delta A}{A_f + \alpha}\ln(1+S_t) + \ln(S_t) \le 0.
\]

Collecting the $\ln(S_t)$ terms:
\[
\frac{\Delta A}{A_f + \alpha}\ln(\tilde{r}_t) + \ln(S_t)\left[\frac{\Delta A}{A_f + \alpha} + 1\right] - \frac{\Delta A}{A_f + \alpha}\ln(1+S_t) \le 0.
\]

Simplifying the coefficient of $\ln(S_t)$:
\[
\frac{\Delta A}{A_f + \alpha}\ln(\tilde{r}_t) + \ln(S_t)\left[\frac{\Delta A + A_f + \alpha}{A_f + \alpha}\right] - \frac{\Delta A}{A_f + \alpha}\ln(1+S_t) \le 0.
\]

Factoring out $\frac{\Delta A}{A_f + \alpha}$:
\[
\frac{\Delta A}{A_f + \alpha}\ln(\tilde{r}_t) + \frac{\Delta A}{A_f + \alpha}\ln(S_t) + \ln(S_t) \le \frac{\Delta A}{A_f + \alpha}\ln(1+S_t).
\]

Grouping terms:
\[
\frac{\Delta A}{A_f + \alpha}[\ln(\tilde{r}_t) + \ln(S_t)] + \ln(S_t) \le \frac{\Delta A}{A_f + \alpha}\ln(1+S_t).
\]

Using logarithm properties:
\[
\ln\left[(\tilde{r}_t S_t)^{\frac{\Delta A}{A_f+\alpha}}\right] + \ln(S_t) \le \ln\left[(1+S_t)^{\frac{\Delta A}{A_f+\alpha}}\right].
\]

Combining logarithms on the left side:
\[
\ln\left[S_t \cdot (\tilde{r}_t S_t)^{\frac{\Delta A}{A_f+\alpha}}\right] \le \ln\left[(1+S_t)^{\frac{\Delta A}{A_f+\alpha}}\right].
\]

Simplifying:
\[
\ln\left[S_t^{1+\frac{\Delta A}{A_f+\alpha}} \tilde{r}_t^{\frac{\Delta A}{A_f+\alpha}}\right] \le \ln\left[(1+S_t)^{\frac{\Delta A}{A_f+\alpha}}\right].
\]

Exponentiating both sides:
\[
S_t^{1+\frac{\Delta A}{A_f+\alpha}} \tilde{r}_t^{\frac{\Delta A}{A_f+\alpha}} \le (1+S_t)^{\frac{\Delta A}{A_f+\alpha}}.
\]

Dividing both sides by $S_t$ and rearranging:
\[
S_t^{\frac{\Delta A}{A_f+\alpha}} \tilde{r}_t^{\frac{\Delta A}{A_f+\alpha}} \le \frac{(1+S_t)^{\frac{\Delta A}{A_f+\alpha}}}{S_t}.
\]

This can be rewritten as:
\[
S_t \left(\tilde{r}_t S_t\right)^{\frac{\Delta A}{A_f+\alpha}} \le (1+S_t)^{\frac{\Delta A}{A_f+\alpha}}.
\]

Or equivalently:
\begin{equation}
S_t \left(\frac{\tilde{r}_t S_t}{1+S_t}\right)^{\frac{\Delta A}{A_f+\alpha}} \le 1. \tag{A6}
\end{equation}

Taking the reciprocal of both sides (and reversing the inequality):
\begin{equation}
\frac{1}{S_t} \left(\frac{1+S_t}{\tilde{r}_t S_t}\right)^{\frac{\Delta A}{A_f+\alpha}} \ge 1. \tag{A7}
\end{equation}

\subsection{Surplus Groom Index}

Motivated by condition (A7), the Surplus Groom Index (SGI) is defined as
\[
\text{SGI}_t \equiv \frac{1}{S_t} \left(\frac{1+S_t}{\tilde{r}_t S_t}\right)^{\frac{\Delta A}{A_f+\alpha}}.
\]

This can also be written as:
\[
\text{SGI}_t = \frac{(1+S_t)^{\frac{\Delta A}{A_f+\alpha}}}{S_t^{1+\frac{\Delta A}{A_f+\alpha}} \tilde{r}_t^{\frac{\Delta A}{A_f+\alpha}}}.
\]

The index provides a measure of marriage market imbalance under monogamy and the maintained demographic structure. When $\text{SGI}_t \ge 1$, there is a surplus of grooms (shortage of brides) in the marriage market. Higher values of the index indicate more severe bride shortages.

\subsection{Interpretation}

The Surplus Groom Index incorporates:
\begin{itemize}
\item $S_t$: The sex ratio at birth (females per male)
\item $\tilde{r}_t$: Effective fertility rate (accounting for child mortality)
\item $\Delta A$: The spousal age gap (male age at marriage minus female age at marriage)
\item $A_f + \alpha$: Female age at marriage plus average time interval from marriage to first birth
\end{itemize}

The interpretation is as follows:
\begin{itemize}
\item $\text{SGI}_t = 1$: Marriage market is exactly balanced
\item $\text{SGI}_t > 1$: Surplus of grooms (shortage of brides). The higher the value, the more severe the shortage
\item $\text{SGI}_t < 1$: Surplus of brides (shortage of grooms)
\end{itemize}

The index is particularly useful for comparing the severity of bride shortages across countries or regions, as it accounts for demographic factors beyond just the sex ratio at birth.

\clearpage

\section{Measurement of Age at Marriage (SMAM)}

Average ages at marriage for men and women are not directly reported in Indian census data for all periods. To construct consistent measures of marriage timing, this paper employs the Singulate Mean Age at Marriage (SMAM) method originally proposed by \citet{hajnal1953}.

SMAM measures the average age at first marriage among those who eventually marry, using cross-sectional information on the proportion never married by age. Let $u(a)$ denote the proportion of individuals who are never married at age $a$, and let $\omega$ denote the upper age limit beyond which first marriage is assumed not to occur. The SMAM is defined as
\[
\text{SMAM} = \omega - \int_0^{\omega} u(a) \, da.
\]
In discrete form, as implemented using census age-group data, the integral is approximated by summing age-specific proportions never married across age intervals. Separate SMAM measures are constructed for men and women, yielding $A_m$ and $A_f$, respectively.

This approach yields internally consistent measures of marriage timing across states and census rounds and is standard in demographic analysis when direct information on age at marriage is unavailable.

\end{document}